# Forecasting Visitor Demand from Open AI-Camera Counts and Weather: Lessons from Japan for Saudi Heritage Sites

Amil Khanzada
Regional Revitalization Headquarters
University of Fukui
Fukui, Japan
amil.k@u-fukui.ac.jp

Mohammed Sohail
College of Engineering
Abu Dhabi University
Abu Dhabi, UAE

Jana Wasfi
Department of Computer Engineering
Taibah University
Madinah, Saudi Arabia

Ahmad Showail
Department of Computer Engineering
Taibah University
Madinah, Saudi Arabia
ashowail@taibahu.edu.sa

*Abstract* - Small heritage and nature sites need a few days' notice of busy days to plan staff, overflow parking and shuttles, but most have neither sensors nor forecasting expertise. We test what one artificial intelligence (AI) camera counter per site, a public holiday calendar and a free weather forecast can predict. Using 22 months of open AI-camera counts from three contrasting sites in Fukui, Japan (a train station, a coastal cliff and a mountain summit park), we measure how weather changes visits, forecast daily counts up to a week ahead with only the weather forecast available at the time, test busy-day warnings tied to a staffing decision, and ask how much history a new counter needs. The more exposed the site, the more weather matters: 10 mm of rain comes with 3% fewer detections at the station, 9% fewer at the cliff and 37% fewer at the summit park. With the forecasts actually issued, three-day-ahead weighted absolute percentage error (WAPE) at the summit park falls from 38.5% to 31.9% (Diebold-Mariano $p<0.001$), and the gain holds to five days; the sheltered station gains nothing. Three-day busy-day warnings with forecast weather reach, on past data and under the primary definitions, an area under the receiver operating characteristic curve (AUC) of 0.87-0.93, and weather lifts the overflow car park warning's precision from 0.60 to 0.82. For a new counter, blending a learned model with last week's same-hour counts cuts hourly error after only three days of training, pending a real installation. All inputs are off the shelf, and we outline how the approach could apply to heritage sites in Saudi Arabia.

## I. Introduction

Smart-city programmes increasingly treat tourism as something to manage, not only promote [1]. Large attractions can afford ticketing systems, dense sensors and bespoke deep-learning forecasters [2]; most smaller heritage and nature sites have none of this, yet face the same question every week: will we need extra staff, the overflow car park or a shuttle in the next few days?

Edge artificial intelligence (AI) cameras now make continuous counting affordable at small sites, and some regions publish the counts as open data. Fukui Prefecture in Japan is one of them, and its heavy winter snow, coastal wind and rain, and hot, humid summers make it a demanding test for weather-aware forecasting. Fukui has ranked first in the Japan Research Institute's happiness ranking of the 47 prefectures in every edition since 2014 [3], yet in 2025 it ranked last of the 47 prefectures for foreign overnight stays and 35th for domestic ones [4]. Since December 2024 it has released five-minute AI-camera counts for three tourism sites: Fukui Station, a bullet train transit hub; Tojinbo, a coastal cliff; and the Rainbow Line, a summit park overlooking the Mikata Five Lakes, reached by a scenic road [5].

So far these data have been used to describe, not to forecast: our Distributed Human Data Engine (DHDE) study [6] estimated same-day arrivals and the visits lost to bad weather, and Bian et al. [7] mapped visitor origins from licence-plate cameras. Forecasting studies at attractions have relied on years of ticket data [8], [2]. Closest to ours, Matsunaga et al. forecast hourly pedestrian counts in Takayama from AI cameras and weather and send a congestion calendar to local shops [9], and Bollenbach et al. predict beach occupancy four hours to three days ahead with weather forecasts [10]. To our knowledge, no study has combined open AI-camera counts, the weather forecasts actually issued, busy-day warnings tied to a staffing decision and a test of how soon a new counter becomes useful.

This paper tests that combination. Our contributions are:

- Weather matters more at exposed sites [11], [12]. Our three sites show the same pattern, and weather forecasts are worth more at the exposed ones too.
- Forecasts up to a week ahead. Each forecast uses only the weather forecast issued at the time, so we can see how many days ahead weather still helps.
- Warnings tied to a staffing decision. Busy-day warnings three days ahead, including one for the Rainbow Line's overflow car park and shuttle.
- Cold start. How much history a new counter needs, and a rule that makes a three-day model useful.

## II. Related Work

*Tourism demand forecasting.* Forecasting reviews and competitions show that simple statistical methods are hard to beat on tourism data, and that a new method must beat a naive baseline on data it has not seen [13], [14]. Studies of single attractions forecast daily visits from search, weather and holiday data [8]; others forecast hourly demand across many attractions with deep learning [2].

*Sensing visitors.* Bian et al. [7] described visitor origins from plate cameras and our DHDE study [6] same-day arrivals from the Fukui cameras; neither forecast days ahead, used weather forecasts, issued warnings or studied new counters.

*Weather and visitation.* How much weather affects visits depends on the place: mobile-phone data show different responses in parks, shopping streets and malls [11], and Madden et al. forecast trail use from people counters and weather [12]. A Dutch beach study used weather forecasts to predict crowds up to 14 days ahead as probabilistic forecasts, from which exceedance probabilities can be read off [15]; we instead train a separate classifier for the busy-day decision, because the operator's threshold is a fixed count and a dedicated model can be judged on precision and recall at it.

*Cold start.* In traffic forecasting, models trained at other locations help new sensors [16]; we deliberately test the harder case of a single counter with no neighbour to borrow from. At the SAIL2025 festival in Amsterdam, with no past data, a model retrained during the event on live sensor and weather feeds gave four-hour crowd forecasts [17].

*Saudi context.* Crowd research in Saudi Arabia centres on Hajj and Umrah: congestion forecasting at Masjid al-Haram [18], surveys of information and communication technology for the pilgrimage [19], epidemic models for mass religious gatherings [20] and AI-enabled Internet of Things (IoT) monitoring of city events [21]. Other Saudi work covers IoT and AI for smart cities [22], [23], a micro-mobility pilot in Riyadh [24] and how visitors respond to digital technology at sacred heritage sites [25]. We found no published work on forecasting visitors at the kingdom's quieter heritage sites, which have few sensors.

## III. Data

### Camera counts

The counts [5] are MIT-licensed open data starting on 20 December 2024. The cameras, documented in [6], are DNN Network Bullet IP Camera U2SM-B units (IP66, outdoor) that run You Only Look Once (YOLO) detection on the device, so only counts leave the camera. Fig. 1 shows the sites, Table I the feeds and Fig. 2 the full series:

- *Fukui Station:* a camera at the tourist information centre by the east entrance counts people.
- *Tojinbo: a shopping-street camera counts people.*
- *Rainbow Line:* licence-plate cameras at the gates of two summit car parks count vehicles, not people. Lot 2 is an overflow car park that opens only on busy days, when the operator must also run a shuttle to the summit and call in extra staff; knowing a few days ahead whether it will open is the decision our warnings target. The gate cameras also count faces (median 0.32 per plate), our secondary check.

TABLE I. Camera feeds (code4fukui, 20 Dec 2024 to 10 Oct 2026). Wkend and Hol.: median on those days divided by the weekday median. RL = Rainbow Line.

| Feed | Counts | Days | Median/day | Wkend | Hol. |
|---|---|---|---|---|---|
| Fukui Station | People | 642 | 8,343 | 1.49 | 1.65 |
| Tojinbo | People | 641 | 8,173 | 2.07 | 2.35 |
| RL lot 1 | Vehicles | 607 | 174 | 1.84 | 2.20 |
| RL lot 2 | Vehicles (overflow) | 607 | 0 | n/a | n/a |

We exclude a day if its total is zero or fewer than 80% of its 288 five-minute bins are present (an outage or a closure), except that zero days at the overflow RL lot 2 are kept when lot 1 is counting. The counts are detections, not unique visitors: at Tojinbo the camera registered about five detections per officially reported visitor in 2025 (3.3 million scaled detections against 651,000 visitors [26]). We therefore treat the counts as an activity index. We ignore the cameras' age and gender estimates (3-4% of detections).

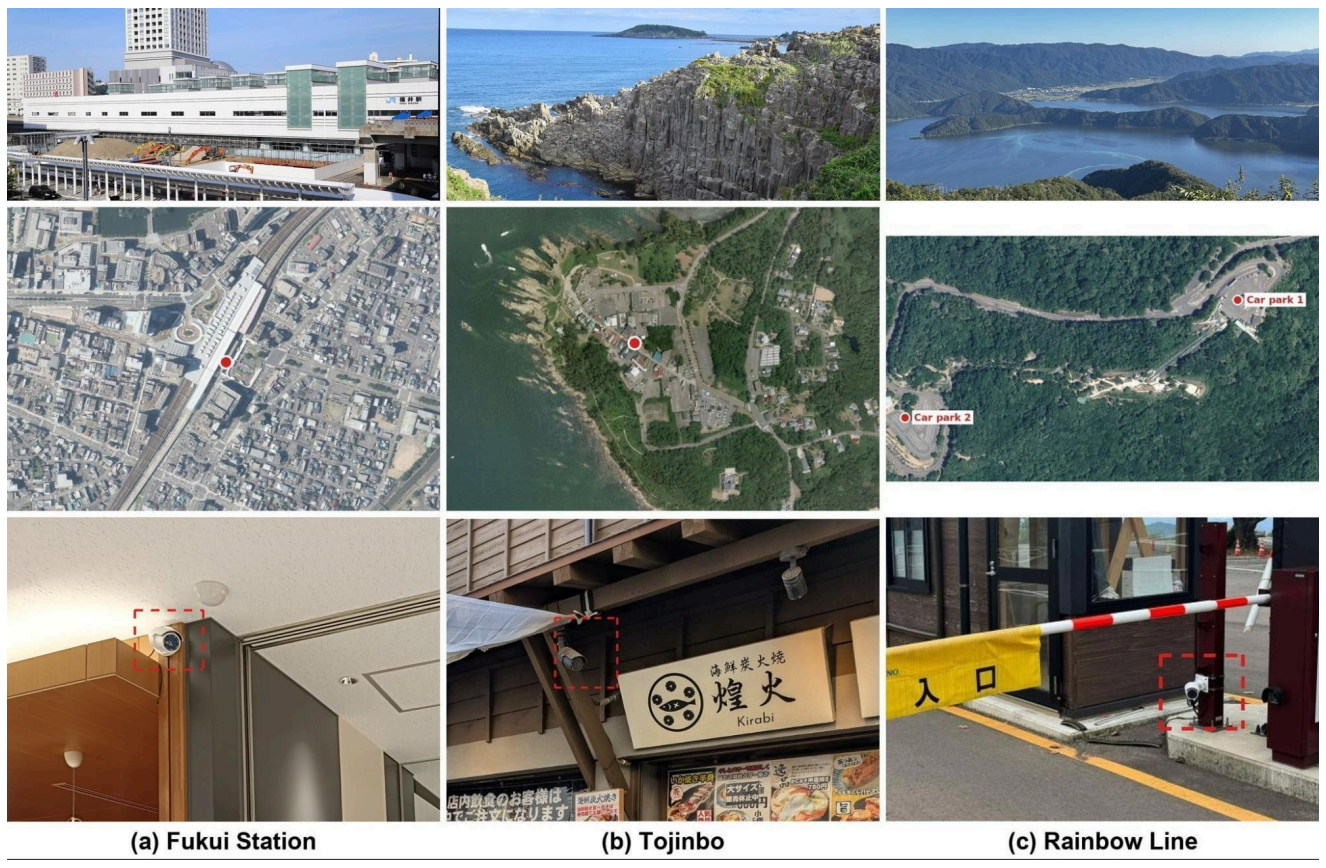

Fig. 1. The three camera sites: the site (top), aerial view (middle) and camera (bottom). (a) Fukui Station tourist information centre; (b) Tojinbo shopping street; (c) Rainbow Line car parks 1 and 2, with the lot 2 gate camera. Aerial imagery: GSI Tiles (Geospatial Information Authority of Japan). Top photos: (a) Araisyohei, CC BY 4.0, and (b) 雷太, CC BY 2.0, via Wikimedia Commons; (c) the authors. Camera photos: (a), (b) the authors; (c) Fukui Prefectural Tourism Federation.

### Weather and holiday calendar

Weather comes from the nearest Japan Meteorological Agency (JMA) stations (Fukui City, Mikuni and Mihama) [27] for the sensitivity analysis, and from Open-Meteo [28], a free global service, for forecasting. Open-Meteo archives the forecasts issued one to seven days before each date (previous-runs application programming interface, API), so we can replay what was known at the time. The daily forecasting models are trained on Open-Meteo's own hourly observation archive (ERA5-based), so the weather they learn from and the forecasts they receive at test time come from the same provider and site coordinates; the cold-start models and Table II use JMA hourly observations. We aggregate rainfall, peak wind gust, sunshine hours and mean temperature over each site’s active hours; for JMA stations, which report no gusts, we use the highest hourly mean wind.

Active hours are those that hold, on average, at least 2% of the day's detections in that calendar month. We set them from pre-test counts only and freeze them before evaluation; November, which has no pre-test data, uses 09:00-17:00, and the overflow lot uses lot 1's hours. Holidays are Japan's statutory public holidays (jpholiday library); a holiday on a weekend counts as a holiday, so the day types are mutually exclusive. Obon and the year-end break are not statutory, so the daily models leave them to the month terms; the cold-start models treat 29 December to 3 January as days off.

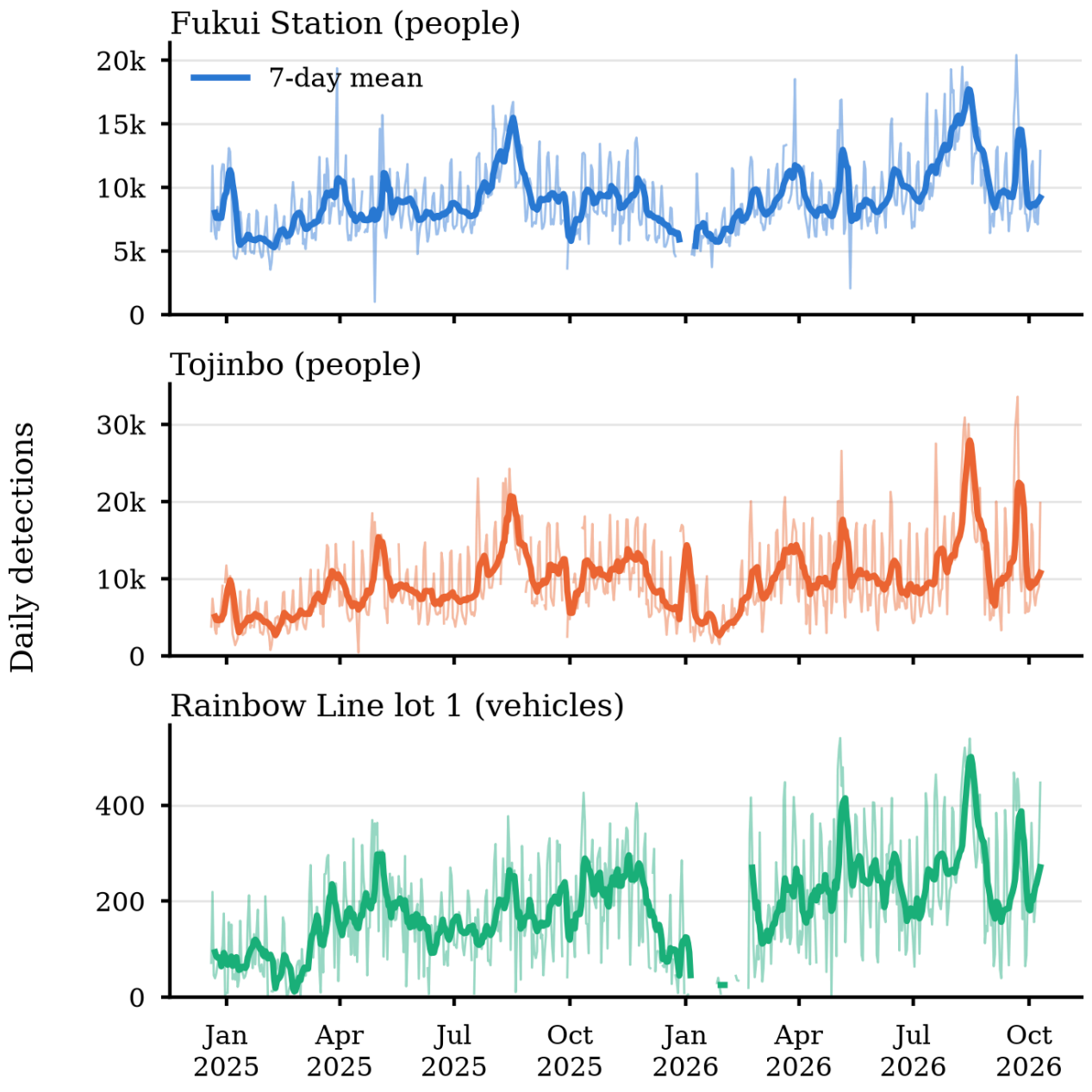

Fig. 2. Daily detections at the three sites, 20 Dec 2024 to 10 Oct 2026 (thin line: daily; thick line: 7-day mean). Gaps are excluded days.

## IV. Methods

### *Weather sensitivity*

To measure how weather moves daily counts, we regress log(1+count) for each feed in Table II on active-hours rainfall, peak wind, sunshine and temperature, controlling for day of week, holidays and year-month. Standard errors are heteroscedasticity- and autocorrelation-consistent (HAC, seven lags). Coefficients $b$, scaled to the weather changes in Table II, are reported as approximate percentage associations with daily counts, $100(\exp(b) - 1)$.

### *Daily forecasts up to a week ahead*

Let $y_d$ be the count on day $d$ and $z_d$=log(1+$y_d$). A forecast issued $h$ days before $d$ ($h$=1,…,7) uses only information available at $d$-$h$. We compare four models:

- *Naive:* the same weekday last week, $\hat{y}_d = y_{d-7}$.
- *Calendar ridge:* ridge regression ($\alpha$=1) on $z_d$ with day-of-week, holiday and month indicators, the 28-day mean of $z$ up to $d$-$h$, and two lags, $z_{d-7}$ and the latest available value $z_{d-h}$.
- *Calendar + forecast weather:* the same model plus active-hours rain, peak wind, sunshine and temperature, and a rain × day-off interaction. It is trained on observed weather, but at test time it receives the Open-Meteo forecast issued $h$ days earlier, so forecast error is included.
- *Gradient boosting: histogram gradient-boosted trees (scikit-learn) on the same inputs, a standard nonlinear check on the linear ridge model.*

As standard reference points we add two classical time-series models, both on log(1+$y$): exponential smoothing (ETS) with weekly seasonality and a damped trend, and a seasonal autoregressive integrated moving-average model with exogenous forecast weather, SARIMAX(1,0,1)(1,0,1)$_7$. Every model is refitted each Monday, so a forecast for day $d$ never sees counts after $d-h$; the fixed settings above were chosen before the test year, which runs from 13 October 2025 to 10 October 2026. As a check against over-tuning, we also tune ridge α and the boosting settings on the first 26 weeks of the test year and score them on the last 26, at leads 1, 3, 5 and 7 for the three sites (12 tests, Holm-corrected). We report weighted absolute percentage error, WAPE = 100 $\Sigma_d\, |y_d - \hat{y}_d| \,/\, \Sigma_d\, y_d$, and the share of days forecast within ±30% of the actual count. Whether one model beats another is tested with the Diebold-Mariano test on absolute errors [29] (Bartlett HAC variance, lag 7). Intervals come from a weekly block bootstrap.

### *Busy-day warnings*

A day counts as busy if it reaches the site's 80th percentile of training counts, recomputed at each weekly refit (a rolling top-20% threshold). Because counts rose during the test year, we also test two alternatives: a threshold frozen before the test year, and one set from past days in the same calendar month. At the Rainbow Line we also predict whether the overflow lot will open (at least 20 plates at lot 2). A logistic regression (L2, C=1) on the same forecasting inputs gives the probability of a busy day three days ahead, and one to seven days ahead in the lead-time comparison. We report the area under the receiver operating characteristic curve (AUC) and precision and recall at a probability of 0.5. Under the same-month definition, test days whose calendar month has no training data are excluded.

### *Cold start for a new counter*

To mimic a newly installed camera we hide all counts before an assumed installation date and train only on the $n$ days that follow it, for $n$ = 3, 7, 14, 28, 56, 91, 182 and 365. We forecast the next day's 24 hourly counts at the two people-counting sites (Fukui Station and Tojinbo) over ten consecutive 28-day windows from 20 December 2025 to 25 September 2026. The windows start a year after the counts begin so that the 365-day history fits; each site uses only its own history, and likely sensor faults are dropped by an hourly outage rule applied before masking. While the week-earlier value is still masked, the baseline falls back to the same hour the day before (then zero); the shrunk model's baseline falls back to the day before and then to the training-period hourly profile, which also fills missing ridge lags. We compare three forecasts: the same hour last week; an hourly ridge model ($\alpha$=3) on the hour × day-off profile, the counts 24 and 168 hours earlier, and observed (JMA) hourly weather; and a shrunk model that blends the two in log space, giving the ridge a weight $w = \min(n/56, 1)$, so it takes over fully at 56 days. We also run the shrunk model with the archived day-ahead forecast at test time, and without weather. Hourly WAPE is pooled over all windows (weekly block bootstrap intervals).

## V. Results

### *Weather sensitivity rises with exposure*

The more exposed the site, the stronger the weather association (Table II): 10 mm of rain comes with about 3% fewer detections at the sheltered station but 37% fewer vehicles at Rainbow Line. Sunshine follows the same order: three more hours of sun come with 17% more vehicles at Rainbow Line and 5% more detections at Tojinbo, but no change at the station. Open-Meteo weather gives the same picture (−1%, −13%, −36% per 10 mm).

TABLE II. Association between active-hours weather and daily counts: % change, with the 95% interval in brackets. Ordinary least squares (OLS) on log(1+count) with weekday, holiday and year-month fixed effects and HAC errors; RL lot 1 counts vehicles.

| Site | 10 mm rain | +5 m/s wind | +3 h sunshine | +5 °C |
|---|---|---|---|---|
| Station | −3 [−6,−1] | +4 [−2,+10] | 0 [−2,+1] | +2 [−1,+5] |
| Tojinbo | −9 [−13,−5] | −10 [−17,−3] | +5 [+3,+8] | +12 [+7,+18] |
| RL lot 1 | −37 [−47,−24] | −13 [−25,+2] | +17 [+12,+22] | +7 [−2,+16] |

### *Forecast accuracy by lead time*

The calendar alone already beats the naive baseline everywhere (Fig. 3). It cuts one-day WAPE by more than a third at the people counters (from 19.3% to 11.9% at Fukui Station) but by less than a fifth at Rainbow Line (41.5% to 34.0%), and error grows with lead time; WAPE and p-values for every model, site and lead are in the repository.

TABLE III. Daily WAPE (%) with the weather forecast actually available at issue time (Open-Meteo previous runs), test 13 Oct 2025 to 10 Oct 2026. Calendar = calendar ridge; + Forecast = the same model with the issued weather forecast; Change = + Forecast minus Calendar with 95% block-bootstrap confidence interval (CI); DM $p$ = unadjusted Diebold-Mariano p-value.

| Site | Lead | Calendar | + Forecast | Change [95% CI] | DM $p$ |
|---|---|---|---|---|---|
| Fukui Station | 1d | 11.88 | 12.04 | +0.2 [−0.1, +0.5] | 0.313 |
| Fukui Station | 3d | 14.43 | 14.67 | +0.2 [0.0, +0.6] | 0.122 |
| Tojinbo | 1d | 23.29 | 21.57 | −1.7 [−3.4, −0.3] | 0.018 |
| Tojinbo | 3d | 27.16 | 26.04 | −1.1 [−3.0, +0.7] | 0.224 |
| Rainbow Line | 1d | 33.95 | 27.38 | −6.6 [−9.5, −3.6] | <0.001 |
| Rainbow Line | 3d | 38.50 | 31.87 | −6.6 [−9.7, −3.6] | <0.001 |

Adding the weather forecast helps exactly where Table II says it should (Table III):

- *Rainbow Line:* the gain is significant up to five days ahead and survives Holm correction at one, three and five days in the tuning check (adjusted $p < 0.01$). Against a perfect-weather reference (the same model given observed weather at test time: WAPE 26.8%, 31.5% and 32.1% at one, three and six days), the issued forecast keeps 92% [66, 113] of the gain one day ahead and 94% [73, 119] three days ahead, but only 23% [−29, 56] at six days. It remains the hardest site: forecasts land within ±30% on 63% of days one day ahead and 54% three days ahead.
- *Tojinbo:* weather is significant one and seven days ahead ($p$ = 0.018 and 0.041), but neither lead is significant in the Holm-corrected tuning check.
- *Fukui Station:* weather adds nothing, but the station is predictable anyway, within ±30% on 90-95% of days at every lead.

Gradient boosting beats forecast-weather ridge at the two exposed sites at most leads (30.0% against 31.9% at Rainbow Line three days ahead) but not at the station. SARIMAX with the same weather comes within about 3 points of ridge, and ETS without weather trails behind.

### *How well do the warnings work?*

Three days ahead, the warnings tell busy days from ordinary ones well at every site (Table IV), and they still do five days ahead (AUC ≥ 0.85) in this retrospective test. Weather adds little at the people counters but cuts false alarms for the Rainbow Line overflow car park (precision 0.60 to 0.82).

With forecast weather, the results hold under three busy-day definitions (AUC 0.82-0.93; code repository), although rising counts make 30-50% of test days busy rather than 20%, and the overflow car park AUC stays 0.88-0.90 however the days on which faces but not plates show lot 2 in use (11 in the scored test year) are treated.

TABLE IV. Busy-day warnings issued three days ahead. Top-20% = busiest fifth of training days; car park 2 in use = at least 20 plates at the overflow lot (Section IV); RL = Rainbow Line.

| Warning | Weather | AUC [95% CI] | Prec. | Rec. |
|---|---|---|---|---|
| Station: top-20% day | none | 0.92 [0.87, 0.96] | 0.78 | 0.62 |
| Station: top-20% day | 3-day forecast | 0.92 [0.87, 0.96] | 0.79 | 0.57 |
| Tojinbo: top-20% day | none | 0.91 [0.85, 0.96] | 0.82 | 0.64 |
| Tojinbo: top-20% day | 3-day forecast | 0.93 [0.88, 0.97] | 0.86 | 0.64 |
| RL: top-20% day | none | 0.81 [0.75, 0.87] | 0.78 | 0.45 |
| RL: top-20% day | 3-day forecast | 0.87 [0.83, 0.91] | 0.84 | 0.54 |
| RL: car park 2 in use | none | 0.87 [0.80, 0.94] | 0.60 | 0.55 |
| RL: car park 2 in use | 3-day forecast | 0.89 [0.82, 0.95] | 0.82 | 0.59 |

Pooling both Rainbow Line gates (604 days), face counts track plates (correlation 0.77) but forecast worse (three-day WAPE 66% against 34%), so plates remain the primary series.

### *How much history does a new counter need?*

A model trained on a few days is worse than repeating last week (Fig. 4): with three days of history, hourly ridge trails last week's same hour by 9 to 14 WAPE points, and it catches up only after about a week at the cliff and four weeks at the station. Blending it with last week's pattern removes the penalty: with three days of history the shrunk model already beats the baseline at both sites, with 95% intervals below zero, though not at every history length (at 56 days the station’s interval is −6.3 to +1.0 points). The gain reaches −10.1 points at the cliff after four weeks and keeps growing over a full year as the model sees more of the seasons. With the archived day-ahead forecast instead of observed weather the gains barely move (−9.7 [−14.7, −5.3] at the cliff after four weeks); weather itself adds about 2 points at the cliff after four weeks and costs 0.6 points at the station.

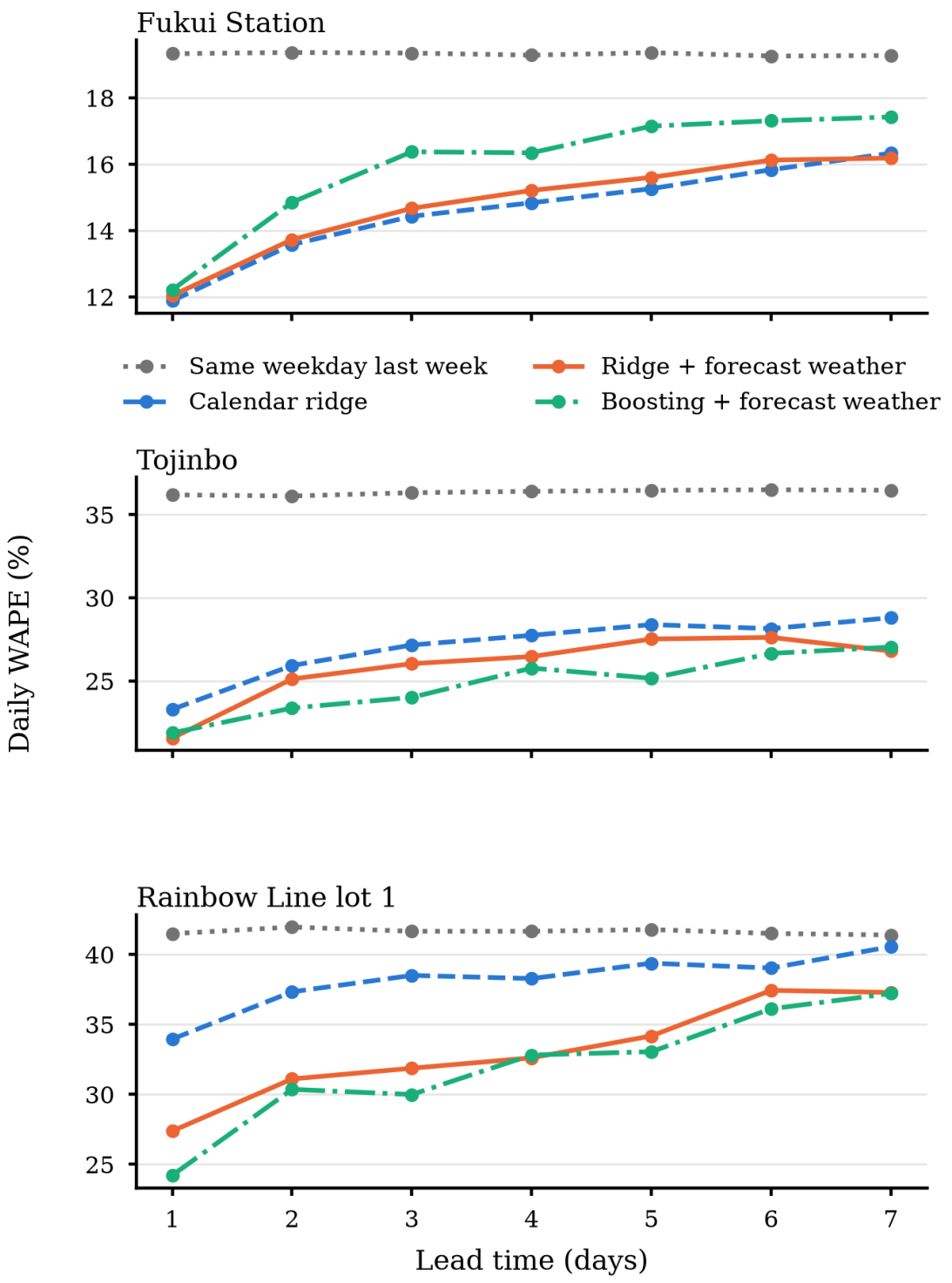

**Fig. 3.** Daily forecast error by lead time over the 12-month test period.

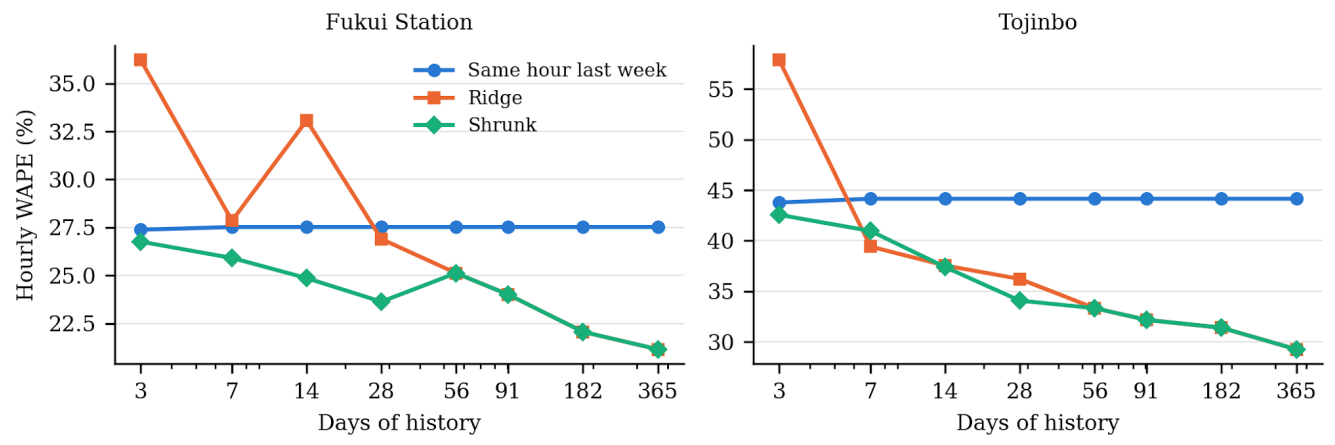

Fig. 4. Cold start: hourly day-ahead error by days of history at the two people counters (Rainbow Line vehicle counts are too sparse for hourly error). “Shrunk” blends ridge with last week's same-hour counts until 56 days.

## VI. Transferability and Future Work

The approach can move to other sites, but counting quality, calendar effects, busy-day definitions, weather sensitivity and the history a new counter needs must be checked locally. We look to Saudi Arabia because it is Fukui's opposite on the surface, hot and dry rather than snowy and wet, yet similar underneath: beyond its famous holy sites, many heritage sites are quiet, little sensed and driven by calendar and weather. Weather forecasts should be tested first at open-air sites such as Mount Uhud in Madinah, not roofed mosques or transit hubs. The pipeline would also need Hijri-period indicators (Ramadan, the Eids), evening active hours, and heat terms in place of rain. In Saudi Arabia the binding weather variable differs by region, from rising summer heat in Madinah [30] to summer rain in the Asir highlands [31], and the calendar follows the Hijri year: domestic tourism peaked during Ramadan and Eid al-Fitr in 2025 [32]. Hegra in AlUla, visited only on capacity-capped guided tours [33], shares the Rainbow Line's staffing decision but not its data: bookings already give a lead signal, so a camera-plus-weather model would complement them rather than replace them. Next steps are a live trial that sends the warnings to an operator and measures whether staffing changes, and a replication comparing a heat-exposed site in Madinah with a rain-exposed one in Asir.

## VII. Limitations and Conclusion

*Limitations.* Three sites over 22 months give only one or two seasonal cycles. The cameras count detections, not unique visitors, and the Rainbow Line counts vehicles rather than people, with little ground truth for either. All results are retrospective: forecasts are replayed from the Open-Meteo archive rather than official JMA forecasts, the busy-day cut-offs are our own, and whether warnings change staffing decisions is untested. The weather models are trained on observed weather and applied to forecasts (perfect prognosis), so noisier long-lead forecasts are not discounted; lead-specific training on archived forecasts is a next step. The 56-day blending ramp was fixed in advance; quieter sites may need a validation-based weight. The cold-start test covers only the two people counters.

*Conclusion.* At three Fukui sites, one AI camera, a holiday calendar and archived weather forecasts support daily forecasts up to a week ahead and flag busy days three days ahead on past data. The weather gains are largest at the summit park, where, with the forecasts issued at the time, they persist to about five days ahead. A new camera can give useful forecasts within days if its model is blended with last week's counts, though this needs testing on a real installation. An off-the-shelf camera and public data make it a low-cost starting point for quieter sites, including Saudi heritage sites.

## Acknowledgment

We thank Code for FUKUI and Fukui Prefecture for publishing the AI-camera data, and the site operators who described their staffing decisions. This activity was supported by the Japan Science and Technology Agency (JST) through the FY2026 International Youth Science Exchange Program (Sakura Science Program), hosted by the University of Fukui.

### Data and code availability

Data, code and a one-command reproduction of every result: https://github.com/amilkh/paper-icaisc-supplement.

### Declaration of AI Use

The authors used generative AI tools (ChatGPT and Claude) for language editing and as coding assistants for the analysis code, figures and repository. The authors designed the study, checked every reported number against the code, and take full responsibility for the paper.